\documentclass{amsart}
\usepackage{amssymb}
\usepackage{latexsym}
\date{\today}

\newcommand{\Z}{{\mathbb Z}}
\newcommand{\R}{{\mathbb R}}
\newcommand{\C}{{\mathbb C}}
\newcommand{\N}{{\mathbb N}}

\newcommand{\Q}{{\mathbb Q}}

\newtheorem{theorem}{Theorem}

\newtheorem{lemma}{Lemma}
\newtheorem{prop}{Proposition}

\renewcommand{\Im}{\mathrm{Im} \, }
\renewcommand{\Re}{\mathrm{Re} \, }

\def\e{\varepsilon}

\begin{document}
\title{Upper Bounds in Quantum Dynamics}

\author{David Damanik}

\address{Mathematics 253--37, California Institute of Technology, Pasadena, CA 91125,
USA}

\email{damanik@caltech.edu}

\author{Serguei Tcheremchantsev}

\address{UMR 6628--MAPMO, Universit\'{e} d'Orl\'{e}ans, B.P.~6759, F-45067 Orl\'{e}ans Cedex,
France}

\email{serguei.tcherem@labomath.univ-orleans.fr}

\begin{abstract}
We develop a general method to bound the spreading of an entire wavepacket under
Schr\"odinger dynamics from above. This method derives upper bounds on time-averaged
moments of the position operator from lower bounds on norms of transfer matrices at
complex energies.

This general result is applied to the Fibonacci operator. We find that at sufficiently
large coupling, all transport exponents take values strictly between zero and one. This
is the first rigorous result on anomalous transport.

For quasi-periodic potentials associated with trigonometric polynomials, we prove that
all lower transport exponents and, under a weak assumption on the frequency, all upper
transport exponents vanish for all phases if the Lyapunov exponent is uniformly bounded
away from zero. By a well-known result of Herman, this assumption always holds at
sufficiently large coupling. For the particular case of the almost Mathieu operator, our
result applies for coupling greater than two.
\end{abstract}

\maketitle

\section{Introduction}

If $H$ is a self-adjoint operator on a separable Hilbert space $\mathcal{H}$, the
time-dependent Schr\"odinger equation of quantum mechanics, $i \partial_t \psi = H \psi$,
leads to a unitary dynamical evolution in $\mathcal{H}$,
\begin{equation}\label{schdyn}
\psi(t) = e^{-itH} \psi(0).
\end{equation}
Of course, the case of main interest is where $\mathcal{H}$ is given by $L^2(\R^d)$ or
$\ell^2(\Z^d)$, $H$ is a Schr\"odinger operator of the form $-\Delta + V$, and $\psi(0)$
is a localized wavepacket.

Under the time evolution \eqref{schdyn}, the wavepacket $\psi(t)$ will in general spread
out with time. Finding quantitative bounds for this spreading in concrete cases and
developing methods for proving such bounds has been the objective of a great number of
papers in recent years.

These questions are particularly complicated and interesting in situations where singular
continuous spectra occur. The discovery of the occurrence of such spectra for many models
of physical relevance has triggered some activity on the quantum dynamical side. For
models with absolutely continuous spectra, on the other hand, quantum dynamical questions
have been investigated much earlier in the context of scattering theory and have been
answered to a great degree of satisfaction. Models with point spectrum or, more
restrictively, models exhibiting Anderson localization, have also been investigated from
a quantum dynamical point of view. While it is not true in general that spectral
localization implies dynamical localization \cite{djls, T3}, it has been shown that the
three main approaches to spectral localization, the fractional moment method
\cite{aenss,am,asfh}, multi-scale analysis \cite{fs,fmss,dk}, and the
Bourgain-Goldstein-Schlag method \cite{bg,goldsch}, yield dynamical localization as a
byproduct \cite{a,bj,ds,gdb,gk}.

Quantum dynamics for Schr\"odinger operators with singular continuous spectra
 is much
less understood. The two primary examples motivated by physics are the Harper model, also
known as the almost Mathieu operator at critical coupling, which describes a Bloch
electron in a constant magnetic field with irrational flux through a unit cell, and the
Fibonacci model, which describes a standard one-dimensional quasicrystal. Both models
have been studied heavily in both the physics and mathematics communities.

The standard approach to dynamical questions is via the spectral theorem. Recall that
each initial vector $\psi(0) = \psi$ has a spectral measure, defined as the unique Borel
measure obeying
$$
\langle \psi, f(H) \psi \rangle = \int_{\sigma(H)} f(E) \, d\mu_\psi(E)
$$
for every measurable function $f$. Here, $\langle \cdot, \cdot \rangle$ denotes the
scalar product in $\mathcal{H}$. A very important discovery of Guarneri \cite{gu,gu2},
which has been extended by other authors \cite{comb,gs,l,KL}, was that suitable
continuity properties of the spectral measure $d\mu_\psi$ imply lower bounds on the
spreading of the wavepacket. Such continuity properties follow from upper bounds on the
measure of intervals, $\mu_\psi ([E -\e, E+\e]), \ E \in \sigma (H), \ \e \to 0$. Later
on, extensions and refinements of Guarneri's method were developed \cite{BGT,GSB,T2},
which allowed the authors of those papers to obtain more general dynamical lower bounds
that take into account the whole statistics of $\mu_\psi ([E-\e, E+\e]), \ E \in \R$. In
particular, lower bounds on $\mu_\psi([E-\e, E+\e])$ for $E$ from some set of positive
Lebesgue measure imply dynamical lower bounds in the case of both singular continuous and
pure point spectrum \cite{GKT}. Better lower bounds can be obtained if one has
information about both the measure of intervals and the growth of the generalized
eigenfunctions $u_\psi (n,E)$ \cite{KL, T2}.

In the case of Schr\"odinger operators in one space dimension, the required information
on the spectral measure (and on generalized eigenfunctions) is intimately related to
properties of solutions to the differential/difference equation $Hu = Eu$
\cite{dkl,DT2,GKT,JL,JL2,T4}. An analysis of these solution has been carried out in a
number of concrete cases \cite{d1,dkl,GKT,JL,JL2,KL}, giving rise to explicit lower
bounds on spreading rates for (generalizations of) the Fibonacci model, sparse
potentials, and (random) decaying potentials. Thus, in one dimension, there is a
three-step procedure for obtaining lower bounds on spreading rates for a given model:
properties of the model $\leadsto$ solution estimates $\leadsto$ bounds on the measure of
intervals $\mu_\psi ([E-\e, E+\e])$ $\leadsto$ dynamical lower bounds.

There is a second approach to dynamical lower bounds in one dimension, which is based on
the Parseval formula,\footnote{The formula \eqref{parsform} was used earlier by Kiselev,
Killip, Last in their study of dynamical upper bounds on the slow part of the wavepacket
\cite{KKL}, which will be discussed momentarily.}
\begin{equation}\label{parsform}
2\pi \int_0^{\infty} e^{-2t/T} | \langle e^{-itH} \delta_1, \delta_n \rangle |^2 \, dt =
\int_{-\infty}^\infty \left|\langle (H - E - \tfrac{i}{T})^{-1} \delta_1, \delta_n
\rangle \right|^2 \, dE.
\end{equation}
This method was developed in \cite{DST,DT,T}. It has the advantage that it gives
dynamical bounds rather directly without any knowledge of the properties of spectral
measure (although additional information on spectral measure allows one to improve the
results \cite{DST,T}). The input required for this method is based on upper bounds for
solutions corresponding to some set of energies, which can be very small (non-empty is
sufficient \cite{DST,DT}). This method is the basis for the results in \cite{dls,JSS}.
Moreover, a combination of this approach with the traditional one (based on the spectral
theorem) leads to optimal dynamical bounds for growing sparse potentials \cite{T} and the
best dynamical lower bounds for the Fibonacci operator known to this date \cite{DT2}.

To a certain extent, there is a fairly good understanding of how to prove dynamical lower
bounds, particularly in one space dimension, as a consequence of the results described
above. The problem of proving upper bounds, on the other hand, is much harder and has
essentially eluded the researchers in this field up to this point. The known lower bounds
for models with singular continuous spectra do not exclude ballistic motion and hence are
not able to distinguish them from models with absolutely continuous spectra. However,
ballistic motion is not expected to occur for the two main models of interest, the Harper
model and the Fibonacci model. While we explain the terminology below, let us mention at
this point that the motion is expected to be almost-diffusive in the Harper model, while
it is expected to be anomalous (i.e., neither localized, nor diffusive, nor ballistic) in
the Fibonacci model. There is a lot of numerical and heuristic evidence for these claims
in the physics literature. To prove such a statement, one needs two-sided estimates for
the dynamical quantities associated with these models.

Proving upper bounds is so hard because one needs to control the entire wavepacket. In
fact, the dynamical lower bounds that are typically established only bound some (fast)
part of the wavepacket from below and this is sufficient for the desired growth of the
standard dynamical quantities. It is of course much easier to prove upper bounds only for
a portion of the wavepacket and Killip, Kiselev and Last succeeded in doing so
\cite{KKL}. Namely, their work provides explicit criteria for upper bounds on the slow
part of the wavepacket in terms of lower bounds on solutions. They applied their general
method to the Fibonacci operator. Their result further supports the conjecture that this
model exhibits anomalous transport.

It is the purpose of this paper to develop a general method connecting solution
properties to dynamical upper bounds and to use this method to prove dynamical upper
bounds for the Fibonacci operator and the almost Mathieu operator.

The general method is based on the Parseval formula \eqref{parsform}, which has already
proved its utility when studying dynamics. We are able to bound the entire wavepacket
from above, provided that suitable lower bounds for solution (or rather transfer matrix)
growth at complex energies are available.

We apply this method to the Fibonacci operator and establish anomalous transport at
sufficiently large coupling. In combination with the known lower bounds, our result shows
that the large coupling asymptotics of the dynamics follows a law predicted by Abe and
Hiramoto in 1987.

Finally, we prove upper bounds for the almost Mathieu operator and extensions thereof at
sufficiently large coupling. It is shown that positive Lyapunov exponents imply a weak
form of dynamical localization in a sense that will be made precise below. This result
complements a stunning observation by Last who proved almost ballistic transport on a
sequence of time scales for the same model in the case of
some Liouville frequencies. Our
result shows that on other sequences of time scales, one has essentially no transport.
Thus, the almost Mathieu operator with a Liouville frequency exhibits quite interesting
dynamics. We also show that the phenomenon discovered by Last is impossible if the
frequency satisfies a weak Brjuno-type condition.

\medskip

\noindent\textit{Acknowledgments.} It is a pleasure to thank Nikolai Makarov and Mihai
Stoiciu for useful conversations.

\section{Description of the Results and Overview}

Consider a discrete one-dimensional Schr\"odinger operator,
\begin{equation}\label{oper}
[H \psi] (n)=\psi(n+1)+\psi(n-1) + V(n)\psi(n),
\end{equation}
on $\ell^2(\Z)$ or $\ell^2(\Z_+)$, where $\Z_+ = \{1,2,\ldots\}$. In the case of
$\ell^2(\Z_+)$, we will work with a Dirichlet boundary condition, $\psi(0) = 0$, but our
results easily extend to all other self-adjoint boundary conditions.

A number of recent papers (e.g., \cite{DST,DT,DT2,JL,JL2,JSS,KKL,T2,T}) were devoted to
proving lower bounds on the spreading of an initially localized wavepacket, say $\psi =
\delta_1$, under the dynamics governed by $H$, typically in situations where the spectral
measure of $\delta_1$ with respect to $H$ is purely singular and sometimes even pure
point.

A standard quantity that is considered to measure the spreading of the wavepacket is the
following: For $p > 0$, define
\begin{equation}\label{mpo}
\langle |X|_{\delta_1}^p \rangle (T) = \sum_n |n|^p a(n,T),
\end{equation}
where
\begin{equation}\label{ant}
a(n,T)=\frac{2}{T} \int_0^{\infty} e^{-2t/T} | \langle e^{-itH} \delta_1, \delta_n
\rangle |^2 \, dt.
\end{equation}
Clearly, the faster $\langle |X|_{\delta_1}^p \rangle (T)$ grows, the faster $e^{-itH}
\delta_1$ spreads out, at least averaged in time. One typically wants to prove power-law
bounds on $\langle |X|_{\delta_1}^p \rangle (T)$ and hence it is natural to define
the following quantity: For $p > 0$, define the \textit{lower transport exponent}
$\beta^-_{\delta_1}(p)$ by
$$
\beta^-_{\delta_1}(p)=\liminf_{T \to \infty} \frac{\log \langle |X|_{\delta_1}^p \rangle
(T) }{p \, \log T}
$$
and the \textit{upper transport exponent} $\beta^+_{\delta_1}(p)$ by
$$
\beta^+_{\delta_1}(p)=\limsup_{T \to \infty} \frac{\log \langle |X|_{\delta_1}^p \rangle
(T) }{p \, \log T}.
$$
Both functions $\beta^\pm_{\delta_1} (p)$ are nondecreasing.

Another way to describe the spreading of the wave-function is in terms of probabilities.
We define time-averaged \textit{outside probabilities} by
$$
P(N,T)=\sum_{|n|>N} a(n,T).
$$
Following \cite{GKT}, for any $\alpha \in [0, +\infty]$ define\footnote{We take $T^\alpha
-2$ so that $P(T^0 -2, T)=1$ for all $T$.}
\begin{equation}\label{smindef}
S^-(\alpha) = - \liminf_{T \to \infty} \frac{\log P(T^\alpha -2, T) }{\log T}
\end{equation}
and
\begin{equation}\label{spludef}
S^+(\alpha) = - \limsup_{T \to \infty} \frac{\log P(T^\alpha -2, T) }{\log T}.
\end{equation}
For every $\alpha$, $0 \le S^+ (\alpha) \le S^- (\alpha) \le \infty$.

These numbers control the power decaying tails of the wavepacket. In particular, the
following critical exponents are of interest:
\begin{align}
\alpha_l^\pm & = \sup \{ \alpha \ge 0  :  S^\pm (\alpha)=0 \}, \label{aldef} \\
\alpha_u^\pm & = \sup \{ \alpha \ge 0  :  S^\pm (\alpha) < \infty \}. \label{audef}
\end{align}
We have that $0 \le \alpha_l^- \le \alpha_u^- \le 1$, $0 \le \alpha_l^+ \le \alpha_u^+
\le 1$, and also that $\alpha_l^- \le \alpha_l^+$, $\alpha_u^- \le \alpha_u^+$. One can
interpret $\alpha_l^\pm$ as the (lower and upper) rates of propagation of the essential
part of the wavepacket, and $\alpha_u^\pm$ as the rates of propagation of the fastest
(polynomially small) part of the wavepacket; compare \cite{GKT}. In particular, if
$\alpha>\alpha_u^+$, then $P(T^\alpha, T)$ goes to $0$ faster than any inverse power of
$T$. Since a ballistic upper bound holds in our case (for any potential $V$), Theorem 4.1
in \cite{GKT} yields
$$
\lim_{p \to 0} \beta_{\delta_1}^\pm (p) = \alpha_l^\pm
$$
and
$$
\lim_{p \to \infty} \beta_{\delta_1}^\pm (p) = \alpha_u^\pm.
$$
In particular, since $\beta^\pm_{\delta_1} (p)$ are nondecreasing, we have that
\begin{equation}\label{betaalphabound}
\beta_{\delta_1}^\pm (p) \le \alpha_u^\pm \quad \text{ for every } p > 0.
\end{equation}

When one wants to bound all these dynamical quantities for specific models, it is useful
to connect them to the qualitative behavior of the solutions of the difference equation
\begin{equation}\label{eve}
u(n+1) + u(n-1) + V(n)u(n) = z u(n)
\end{equation}
since there are effective methods for studying these solutions. Presently, the known
general results are limited to one-sided estimates of the transport exponents. Namely, as
already alluded to in the introduction, a number of approaches to lower bounds on
$\beta^\pm_{\delta_1}(p)$ have been found in recent years.

It should be stressed that there were no general methods known to bound $\alpha_l^\pm,
\alpha_u^\pm$, or $\beta^\pm_{\delta_1}(p)$ non-trivially from above. In the present
paper we propose a first general approach to proving upper bounds on $\alpha_u^\pm$
(which in turn bound $\alpha_l^\pm$ and $\beta^\pm (p)$ for all $p>0$ from above, as
well). This approach relates the dynamical quantities introduced above to the behavior of
the solutions to the difference equation \eqref{eve} for complex energies $z$. To state
this result, let us recall the reformulation of \eqref{eve} in terms of transfer
matrices. These matrices are uniquely determined by the requirement that
$$
\left( \begin{array}{c} u(n+1) \\ u(n) \end{array} \right) = \Phi(n,z) \left(
\begin{array}{c} u(1) \\ u(0) \end{array} \right)
$$
for every solution $u$ of \eqref{eve}. Consequently,
\begin{equation}
\Phi(n,z) = \begin{cases} T(n,z) \cdots T(1,z) & n \ge 1, \\ \mathrm{Id} & n = 0,
\\ [T(n+1,z)]^{-1} \cdots [T(0,z)]^{-1} & n \le -1, \end{cases}
\label{Phi}
\end{equation}
where
$$
T(m,z) = \left( \begin{array}{cr} z - V(m) & -1 \\ 1 & 0 \end{array} \right).
$$
We have the following result:

\begin{theorem}\label{main1}
Suppose $H$ is given by \eqref{oper}, where $V$ is a bounded real-valued function, and $K
\ge 4$ is such that $\sigma (H) \subseteq [-K+1,K-1]$. Suppose that, for some $C \in
(0,\infty)$ and $\alpha \in (0,1)$, we have
\begin{equation}\label{assumeright}
\int_{-K}^K \left( \max_{1 \le n \le C T^\alpha} \left\| \Phi \left( n,E+ \tfrac{i}{T}
\right) \right\|^2 \right)^{-1} dE = O(T^{-m})
\end{equation}
and
\begin{equation}\label{assumeleft}
\int_{-K}^K \left( \max_{1 \le -n \le C T^\alpha} \left\| \Phi \left( n,E+ \tfrac{i}{T}
\right) \right\|^2 \right)^{-1} dE = O(T^{-m})
\end{equation}
for every $m \ge 1$. Then
\begin{equation}\label{apubound}
\alpha_u^+ \le \alpha.
\end{equation}
In particular,
\begin{equation}\label{bppbound}
\beta^+ (p) \le \alpha \quad \text{ for every } p > 0.
\end{equation}
\end{theorem}

\noindent\textit{Remarks.} (a) If the conditions of the theorem are fulfilled for some
sequence of times $T_k \to \infty$, we get an upper bound for $\alpha_u^-$.\\[1mm]
(b) The statement of the theorem follows from upper bounds for outside probabilities
described in Theorem~\ref{main} below.

\medskip

The next issue we want to address is the stability of the transport exponents
$\alpha_u^\pm$ with respect to suitable perturbations of the potential. For example, it
is reasonable to expect that they are invariant with respect to finitely supported
perturbations. However, such a property has not been established yet. Moreover, the
approaches to lower bounds for transport exponents that are based on dimensional
properties of spectral measures are not suited to prove this invariance. It is well known
that these dimensions are rather sensitive with respect to finite rank perturbations. As
pointed out in \cite{DST,DT}, the approach developed in those papers is stable in the
sense that if it can be applied to a certain model, it can also be applied (and yields
the same bounds) to finitely supported (or even suitable power-decaying) perturbations.

Of course, it is even more desirable to prove that the transport exponents (as opposed to
one-sided bounds) are invariants. The following result establishes this fact:

\begin{theorem}\label{stab}
Let $H_1, H_2$ be two operators of the form \eqref{oper} with bounded potentials and let
$K \ge 4$ be such that $\sigma (H_{1,2}) \subseteq [-K+1, K-1]$. Denote the corresponding
transfer matrices by $\Phi_1, \, \Phi_2$ and the corresponding transport exponents by
$\alpha_{1,u}^\pm, \, \alpha_{2,u}^\pm$. Assume there exists $A>0$ such that for  all $E
\in [-K,K], \ 0<\e \le 1$, $|n| \le 1/\e$,
\begin{equation}\label{h16}
\e^A \|\Phi_1 (n, E+i\e)\| \lesssim \|\Phi_2 (n, E+i\e)\| \lesssim \e^{-A}  \|\Phi_1 (n,
E+i\e)\|.
\end{equation}
Then, $\alpha_{1,u}^\pm = \alpha_{2,u}^\pm$.
\end{theorem}

The bounds \eqref{h16} clearly hold if the potentials differ only on a finite set. Thus,
Theorem~\ref{stab} yields the invariance mentioned above. For example, assume that
dynamical localization holds for the operator $H_1$. That means, in particular, that
$\beta_{1, \delta_1}^+(p)=0$ for every $p>0$ and thus $\alpha_{1,u}^+=0$.
Theorem~\ref{stab} yields $\alpha_{2,u}^+=0$, so that $\beta_{2, \delta_1}^+ (p)=0$ ---
some weak form of dynamical localization for $H_2$. This result should be compared with
that of \cite{djls} about semi-stability of dynamical localization.

\bigskip

In most cases where nontrivial dynamical lower bounds have been proven, the results
obtained imply $\alpha_l^\pm >0$ and $\alpha_u^\pm=1$. This is the case for sparse
potentials \cite{GKT,T}, random decaying potentials \cite{GKT}, the Thue-Morse
Hamiltonian \cite{DT}, and random polymer models \cite{JSS}. Although it has not yet been
proven, it is expected that $\alpha_l^\pm <1$ for some of these models (for sparse
potentials, however, it was shown in \cite{cm} that $\alpha_l^+=1$).

On the other hand, there are models for which it is expected that $\alpha_u^\pm$  may
take values other than zero (this is the case when one has dynamical localization) or one
(ballistic transport at least for $p \gg 1$). Such a situation is called
\textit{anomalous transport} by some authors. More restrictively, $\alpha_u^\pm = 1/2$ is
sometimes called \textit{diffusive transport} and anomalous transport refers to a
situation where one does not have ballistic transport, diffusive transport, or dynamical
localization.

Two very prominent examples are given by the Fibonacci operator and the almost Mathieu
operator. We will apply our general result, Theorem~\ref{main1}, to both of them.

The Fibonacci operator has been studied for decades by mathematicians and physicists. The
potential is given by
\begin{equation}\label{fibpot}
V(n) = \lambda \chi_{[1- \theta,1)}(n\theta \mod 1), \quad \theta = \frac{\sqrt{5}-1}{2}.
\end{equation}
This potential belongs to the more general class of Sturmian potentials, given by
\begin{equation}\label{sturm}
V(n) = \lambda \chi_{[1- \theta,1)}(n\theta + \omega \mod 1)
\end{equation}
with general irrational $\theta \in (0,1)$ and arbitrary $\omega \in [0,1)$. These
sequences provide standard models for one-dimensional quasicrystals.\footnote{See
\cite{sbgc} for the discovery of quasicrystals and \cite{bamo} for surveys on the
mathematical theory of quasicrystals in general and the role of the Fibonacci operator in
particular.} Early numerical and heuristic studies of the spectral properties of the
Fibonacci model were performed by Kohmoto, Kadanoff, Tang \cite{kkt} and Ostlund, Pandit,
Rand, Schellnhuber, Siggia \cite{oprss}. It was suggested that the spectrum is always
purely singular continuous. This was rigorously established by S\"ut\H{o} for the
Fibonacci case \cite{su,su2} and by Bellissard, Iochum, Scoppola, Testard \cite{bist} and
Damanik, Killip, Lenz \cite{dkl} in the general Sturmian case. Abe and Hiramoto studied
the transport exponents for the Fibonacci model numerically \cite{ah,ah1}. They found
that they are decreasing in $\lambda$ and behave like
\begin{equation}\label{fibdynexp}
\alpha_l^\pm , \, \alpha_u^\pm  \asymp \frac{1}{\log \lambda}
\end{equation}
as $\lambda \to \infty$. (Here and in the following, we write $f \asymp g$ if $C^{-1} g
\le f \le C g$ for some $C \ge 1$.) See also \cite{gkp,kkkg} for other numerical studies
of Fibonacci quantum dynamics.

The general approaches to lower bounds for the transport exponent described above have
all been applied to the Fibonacci Hamiltonian (and some Sturmian models). The best lower
bound for $\alpha_l^-$ was obtained by Killip, Kiselev, and Last in \cite{KKL}. It
reads\footnote{The expression for $\zeta(\lambda)$ is given in terms of its large
$\lambda$ asymptotics which is of main interest here. See \cite{DT,KKL} for its values at
small $\lambda$.}
$$
\alpha_l^- \ge \frac{2 \kappa}{\zeta (\lambda) + \kappa+1/2},
$$
where
$$
\kappa =  \frac{\log \frac{\sqrt{17} } {4} } {5 \log \left( \frac{\sqrt{5} + 1}{2}
\right)} \approx 0.0126
$$
and $\zeta (\lambda)$, chosen so that one can prove a result like
$$
\sum_{n = 1}^L \|\Phi (n, E)\|^2 \le C L^{2 \zeta (\lambda) +1}
$$
for energies in the spectrum of $H$ (our definition differs from that of \cite{KKL}),
obeys
$$
\zeta (\lambda) = \frac{3 \log \sqrt{5}}{\log \left( \frac{\sqrt{5} + 1}{2} \right)}
\left( \log \lambda + O(1) \right).
$$
This shows in particular that $\alpha_l^-$ admits a lower bound of the type
\eqref{fibdynexp}.

The best lower bound for $\alpha_u^-$ was found in \cite{DT}, where it was shown that
$$
\alpha_u^- \ge \frac{1}{\zeta (\lambda) + 1}.
$$
In terms of the exponents $\beta_{\delta_1}^-(p)$, the best known lower bounds are (see
\cite{DT2})
$$
\beta^-_{\delta_1}(p) \ge \begin{cases} \frac{p+2\kappa}{(p+1)
(\zeta (\lambda) + \kappa+1/2)}
& p \le 2 \zeta (\lambda)+1, \\
\frac{1}{\zeta (\lambda) + 1} & p>2 \zeta (\lambda) +1. \end{cases}
$$

We also want to mention work on upper bounds for the slow part of the wavepacket by
Killip, Kiselev, and Last, \cite{KKL}. More precisely, they showed that there exists a
$\delta \in (0,1)$ such that for $\lambda$ large enough (so that $p(\lambda)$ defined
in \eqref{plambda} below is less than one),
$$
P(C_2 T^{p (\lambda)}, T) \le 1 - \delta.
$$
Here,
\begin{equation}\label{plambda}
p(\lambda) = \frac{6  \log  \frac{\sqrt{5} + 1}{2} } {\log \xi (\lambda)}
\end{equation}
and
\begin{equation}\label{xilambda}
\xi(\lambda) =  \frac{\lambda - 4 + \sqrt{(\lambda - 4)^2 -12}}{2}.
\end{equation}
See \cite[Theorem~1.6.(i)]{KKL}. However, this result does not say anything about the
fast part of the wavepacket, and in particular, no statement for any of the transport
exponents can be derived.

With the help of Theorem~\ref{main1} we can prove upper bounds for $\alpha_u^+$ for the
Fibonacci model at sufficiently large coupling. These upper bounds show that
\eqref{fibdynexp} is indeed true.

The precise result is as follows:

\begin{theorem}\label{main2}
Consider the Fibonacci Hamiltonian, that is, the operator \eqref{oper} with potential
\eqref{fibpot}. Assume that $\lambda \ge 8$ and let
$$
\alpha(\lambda) = \frac{2 \log \frac{\sqrt{5}+1}{2}}{\log \xi(\lambda)}.
$$
with $\xi(\lambda)$ as in \eqref{xilambda}. Then,
\begin{equation}\label{fibconcl}
\alpha_u^+ \le \alpha (\lambda),
\end{equation}
and hence
\begin{equation}\label{fibconcl2}
\beta^+ (p) \le \alpha (\lambda) \quad \text{ for every } p > 0.
\end{equation}
\end{theorem}
One can observe that $\alpha (\lambda)<p(\lambda)$, where $p(\lambda)$ is the number
given in \eqref{plambda}---the power for which \cite{KKL} proved upper bounds for the
slow part. Note that $\xi(\lambda) = \lambda + O(1)$ as $\lambda \to \infty$. Moreover,
$$
\alpha(8) = \frac{2 \log \frac{\sqrt{5}+1}{2}}{\log 3} \approx 0.876
$$
and $\alpha(\lambda)$ is a decreasing function of $\lambda$ for\footnote{The minimum
value, $\lambda_0$, can be chosen slightly smaller than $8$. By monotonicity, all we need
to require from $\lambda_0$ is that $\xi(\lambda_0) > 1$ and $\alpha(\lambda_0) < 1$.}
$\lambda \ge 8$. Thus, we establish anomalous transport for the Fibonacci Hamiltonian
with coupling $\lambda \ge 8$ and confirm the asymptotic dependence of the transport
exponents $\alpha_u^\pm$ on the coupling constant $\lambda$ that was predicted by Abe and
Hiramoto. We emphasize again that this is the first model for which anomalous transport,
in the sense that $0 < \alpha_l^- \le \alpha_u^+ < 1$, can be shown rigorously.

\bigskip

The almost Mathieu operator is given by \eqref{oper} with potential
\begin{equation}\label{amo}
V(n) = \lambda \cos (2\pi (n \theta + \omega))
\end{equation}
with $\theta$ irrational and $\omega \in [0,1)$. The almost Mathieu operator in the
critical case $\lambda = 2$ is also called the Harper model,
\begin{equation}\label{harper}
V(n) = 2 \cos (2\pi (n \theta + \omega))
\end{equation}
Operators with potential \eqref{amo} have been studied extensively by mathematicians and
physicists for decades. The spectral theory is almost completely understood. For almost
every frequency $\theta$ and phase $\omega$, one has purely absolutely continuous
spectrum for $0 < \lambda < 2$, purely singular continuous spectrum for $\lambda = 2$,
and pure point spectrum with exponentially decaying eigenfunctions (\textit{Anderson
localization}) for $\lambda > 2$.

On the other hand, for $\theta$'s that are very well approximated by rational numbers,
one can use Gordon's Lemma \cite{g} to prove absence of eigenvalues so that localization
fails for these $\theta$'s and all $\omega$'s when $\lambda > 2$. But even for a
Diophantine $\theta$, absence of eigenvalues holds generically, so that for a dense
$G_\delta$ set of $\omega$'s, one has again no eigenvalues \cite{js}. To a certain
extent, the situation was mollified by Jitomirskaya and Last \cite{JL2}, who proved that
for $\lambda > 2$, all irrational frequencies and all phases, the spectral measures are
zero-dimensional.\footnote{This was proved earlier in the special case of Liouville
frequencies by Last \cite{l}.}

On the dynamical side, the result of Last that for $0< \lambda < 2$, every irrational
$\theta$, and every\footnote{Last proves this for almost every $\omega$; it holds for
every $\omega$ by a result of Last and Simon \cite{ls}.} $\omega$, there exists some
absolutely continuous spectrum \cite{l2}, implies ballistic motion in this coupling
regime. In the regime $\lambda > 2$, it is of course desirable to establish dynamical
localization whenever spectral localization holds. For results in this direction, see
\cite{germ,gj,jl3}. As a result, dynamical localization holds for $\lambda
> 2$, Diophantine $\theta$, and almost every $\omega$.

On the other hand, it was not clear what should be expected for the dynamical quantities
$\beta^-_{\delta_1}(p) = \beta^-_{\delta_1} (p, \lambda, \theta, \omega)$ in cases of
exceptional frequencies and/or phases at coupling $\lambda > 2$. Since the
Guarneri-Combes-Last bound is one-sided, zero-dimensionality of spectral measures has no
dynamical implications. In fact, Last proved the following result in \cite{l}:
$\beta^+_{\delta_1} (2, \lambda, \theta, \omega) = 1$ (and hence
$\alpha_u^+(\lambda,\theta,\omega)=1$) for every $\lambda$, $\omega$, and a suitable
$(\lambda,\omega)$-dependent Liouville frequency $\theta$. Del Rio, Jitomirskaya, Last,
and Simon constructed $\theta$ such that $\beta_{\delta_1}^+ (p, 3, \theta, \omega)=1$
for every $p>0, \omega \in [0,1)$ \cite{djls} (their proof given for $p=2$ can be easily
generalized to any $p>0$). Thus, for such a value of $\theta$, \ $\alpha_l^+(3, \theta,
\omega) =1$. Later on, it was shown in \cite{GKT} that given $\lambda>2$, one has
$\alpha_l^+(\lambda, \theta, \omega)=1$ for all $\theta$'s from some dense $G_\delta$
set.

Thus, in these exceptional cases, one can only hope to bound $\alpha_u^- (\lambda,
\theta, \omega)$ from above non-trivially.

Recall the continued fraction expansion of $\theta$,
$$
\theta = a_0 + \cfrac{1}{a_1+ \cfrac{1}{a_2+ \cfrac{1}{a_3 + \cdots}}},
$$
with uniquely determined integers $a_0 \in \Z$ and $a_n \ge 1$, $n \ge 1$; see
\cite{khin} for background information. Truncation of this expansion gives rise to best
rational approximants $p_k/q_k$ of $\theta$,
$$
\frac{p_k}{q_k} = a_0 + \cfrac{1}{a_1+ \cfrac{1}{a_2 + \cdots + \cfrac{1}{a_k}}}.
$$
The following condition on $\theta$ will be useful below:
\begin{equation}\label{weakbrjuno}
\lim_{k \to \infty} \frac{\log  q_{k+1}}{q_k} = 0.
\end{equation}
This condition is satisfied for Lebesgue almost every $\theta$; see, for example,
\cite[Theorem~31]{khin}. It is related to, but much weaker than, the Brjuno condition,
which states that $(\log  q_{k+1})/q_k \in \ell^1$. The Brjuno condition is crucial, for
example, in Yoccoz' work on analytic linearization of circle diffeomorphisms \cite{y}.
Every Diophantine number satisfies the Brjuno condition and the inclusion is strict.

We have the following dynamical results for the almost Mathieu operator at super-critical
coupling:

\begin{theorem}\label{amocoro}
For the almost Mathieu operator with coupling $\lambda > 2$, any irrational $\theta$, and
any $\omega \in [0,1)$, we have $\alpha_u^-(\lambda,\theta,\omega) = 0$ {\rm (}and hence
$\beta^-_{\delta_1}(p,\lambda,\theta,\omega) = 0$ for every $p > 0${\rm )}. If in
addition $\theta$ obeys \eqref{weakbrjuno}, then $\alpha_u^+(\lambda, \theta, \omega)=0$
for every $\omega \in [0,1)$ {\rm (}and hence
$\beta^+_{\delta_1}(p,\lambda,\theta,\omega) = 0$ for every $p > 0${\rm )}.
\end{theorem}

\noindent\textit{Remarks.} (i) In particular, for any given coupling $\lambda > 2$, any
phase $\omega$, and an associated Liouville frequency $\theta$, there is a sequence of
time scales, for which one has almost ballistic transport, but there is essentially no
transport along certain other sequences of time scales. Thus, the Liouville case may
exhibit interesting dynamical phenomena.\\
(ii) Under the rather weak condition \eqref{weakbrjuno} we get a certain form of
dynamical localization. While this result is slightly weaker than the one established by
Germinet and Jitomirskaya under a Diophantine assumption on $\theta$ \cite{gj}, it has
several advantages:
\begin{itemize}

\item The assumption \eqref{weakbrjuno} is much weaker than the one from \cite{gj};
compare \cite[Theorems~9 and 13]{khin}. As mentioned above, any standard Diophantine
condition implies the Brjuno condition, which in turn implies \eqref{weakbrjuno}, and
both inclusions are strict.

\item Our result holds for all phases $\omega$, while Germinet and Jitomirskaya consider
a phase average and, in particular, lose control over zero measure sets. As mentioned
above, even under a Diophantine condition, there are exceptional values of $\omega$ for
which one has no eigenvalues. Germinet and Jitomirskaya cannot make any statement for
these $\omega$'s, while we can handle every $\omega$. In a way, our result bears some
similarity with a dynamical upper bound established by del Rio, Jitomirskaya, Last, Simon
in \cite[Theorem~8.1]{djls} who considered exceptional couplings in a rank-one
perturbation situation: While there are exceptional phases, for which spectral
localization fails, the failure of (the strong form of) dynamical localization cannot be
too severe.

\item Our proof is much shorter. It only relies on positive Lyapunov exponents and does
not need to establish spectral localization first. Note that \cite{gj} relies on
\cite{bg,goldsch,jito}, while our proof is self-contained up to a proof of positive
Lyapunov exponents, which can be proved in a few lines \cite{h}; see also
\cite[pp.~199--200]{cfks}.

\end{itemize}

\medskip

We will prove Theorem~\ref{amocoro} in a more general form. Let us consider an operator
$H = H_{\theta,\omega}$ of the form \eqref{oper} whose potential is given by
$$
V(n) = V_{\theta,\omega}(n) = f(n\theta + \omega),
$$
where $f$ is a trigonometric polynomial, $\theta$ is irrational, and $\omega \in [0,1)$.

The Lyapunov exponent, whose existence follows from Kingman's subadditive ergodic
theorem, is given by
\begin{equation}\label{lyapex}
\gamma_\theta(z) = \lim_{n \to \infty} \frac{1}{n} \int_0^1 \log \|
\Phi_n(z,\theta,\omega) \| \, d\omega = \inf_{n \ge 1} \frac{1}{n} \int_0^1 \log \|
\Phi_n(z,\theta,\omega) \| \, d\omega.
\end{equation}
One gets the same quantity if one considers $n \le -1$.

\begin{theorem}\label{amothm}
Let $H_{\theta,\omega}$ be as just described. Assume that there is a number $\Gamma > 0$
such that $\gamma_\theta(z) \ge \Gamma$ for every $z \in \C$. Then
$\alpha_u^-(\theta,\omega) = 0$ for every $\omega \in [0,1)$. If in addition $\theta$
obeys \eqref{weakbrjuno}, then $\alpha_u^+(\theta, \omega)=0$ for every $\omega \in
[0,1)$.
\end{theorem}

A classical result of Herman \cite{h} states that if $f(\omega) = \lambda \cos (2 \pi
\omega)$, then
$$
\gamma_\theta(z) \ge \log \frac{|\lambda|}{2} \quad \text{ for every } \theta \in \R
\setminus \Q , \; z \in \C.
$$
Thus, Theorem~\ref{amothm} implies Theorem~\ref{amocoro}.

For a general non-constant trigonometric polynomial $f$, Herman's argument shows that the
assumption of Theorem~\ref{amothm} holds for sufficiently large leading coefficient.
Thus, we may state the following

\begin{theorem}\label{analcoro}
Suppose $g$ is a non-constant trigonometric polynomial and $f = \lambda g$. Then, there
exists $\lambda_0$ such that for every $\lambda > \lambda_0$, every irrational $\theta$,
and every $\omega \in [0,1)$, we have $\alpha_u^-(\lambda,\theta,\omega) = 0$. Moreover,
if $\theta$ obeys \eqref{weakbrjuno}, then $\alpha_u^+(\lambda, \theta, \omega)=0$ for
every $\lambda > \lambda_0$ and $\omega \in [0,1)$.
\end{theorem}

Note that Theorems~\ref{amocoro} and \ref{analcoro}, combined with the
Guarneri-Combes-Last bound, provide a new proof of the zero-dimensionality of spectral
measures since we do not use the Jitomirskaya-Last inequality from \cite{JL} on which the
proof in \cite{JL2} was based.

We believe that one can extend Theorem~\ref{amothm}, and hence Theorem~\ref{analcoro}, to
the case of $f = \lambda g$ with $g$ analytic and $\lambda$ sufficiently large.
Positivity of the Lyapunov exponent was shown in this case by Sorets and Spencer
\cite{ss}; see also \cite{bg,gs}. Zero-dimensionality of spectral measures for analytic
potentials and large coupling is a result of Jitomirskaya and Landrigan \cite{jland}.

Given Germinet-Jitomirskaya \cite{gj}, Last \cite{l}, and Theorem~\ref{amocoro} above, we
have a good understanding of the dynamical picture in the super-critical case, $\lambda >
2$. Due to the existence of absolutely continuous spectrum in the sub-critical case
\cite{l2}, $0 < \lambda < 2$, this leaves the intriguing problem of studying the critical
case, $\lambda = 2$, that is, when the potential is given by \eqref{harper}. There are no
rigorous results to this date. Based on numerics \cite{gkp,ah2,kkkg,wa}, it is expected
that one has almost diffusive transport in the sense that the transport exponents are
close, but not equal, to $1/2$.

The organization is as follows. In Section~3 we prove upper bounds for outside
probabilities in terms of the norms of transfer matrices at complex energies; see
Theorem~\ref{main}. These bounds immediately imply Theorem~\ref{main1} above. We also
express the numbers $\alpha_u^\pm$ in terms of transfer matrix norms (see
Theorem~\ref{I}) and prove Theorem~\ref{stab}. In Section~4 we study the Fibonacci
Hamiltonian. Suitable bounds on transfer matrix norms are obtained by studying the traces
of these matrices. Since they obey recursive relations, we have to investigate a
dynamical system, the so-called \textit{trace map}. As opposed to earlier studies of the
trace map, we need to regard this as a complex dynamical system and use ideas from
conformal mapping theory, the Koebe Distortion Theorem in particular, to prove the
required estimates. Once these estimates have been established, Theorem~\ref{main2}
follows from Theorem~\ref{main1}. Theorem~\ref{amothm} is proved in Section~5. Finally,
we make some concluding remarks in Section~6.

\section{Upper Bounds on the Tail of the Wavepacket}

In this section we prove an upper bound on the tail of the (time-averaged) wavepacket in
terms of transfer matrix norms for complex energies. It is shown how this leads to upper
bounds on transport exponents in suitable cases.

Observe that, by unitarity, $\sum_n a(n,T)=1$ for any $T$. For an integer $N \ge 1$,
define
$$
P_r(N,T)=\sum_{n>N} a(n,T) \quad \text{ and }  \quad P_l(N,T)=\sum_{n<-N} a(n,T)
$$
so that $P(N,T) = P_l(N,T) + P_r(N,T)$. We call $P_r(N,T)$ (resp., $P_l(N,T)$) the right
(resp., left) outside probability. Our first goal in this section is to find upper bounds
for $P_r(N,T)$ and $P_l(N,T)$. We will give the details for $P_r(N,T)$. Analogous bounds
for $P_l(N,T)$ can be obtained in the exact same way.

To be precise, we want to find $N(T) \to \infty$ such that
\begin{equation}\label{g1}
\lim_{T \to \infty} P_r(N(T), T)=0.
\end{equation}
The intuitive meaning of \eqref{g1} is that for $T$ large, the wavepacket $\psi(t) =
e^{-itH} \delta_1$, $0 \le t \le T$, has essentially no weight in the region $(N(T),
\infty)$ for times up to $T$, at least in a time-averaged sense. A ballistic upper bound
shows that \eqref{g1} always holds if we let $N(T)=T^{1+\nu}$ for some $\nu>0$. In the
case of a dynamically localized system (e.g., when the potential is given by independent
identically distributed random variables), \eqref{g1} trivially holds for any $N(T)$ such
that $N(T) \to \infty$. The most interesting case is the intermediate one, where
\eqref{g1} holds for $N(T)=T^\alpha$ with some $\alpha \in (0,1)$ (and does not hold for
$N(T)=T^{\gamma}$ with some $0<\gamma<\alpha$).

When proving \eqref{g1} and the analogous bound for $P_l(N(T),T)$, two kinds of results
are of interest:\\[2mm] (i) Find $N(T)$ such that $P_r(N(T),T)$ and $P_l(N(T),T)$ go to
zero fast; for example, faster than any inverse power of $T$.
In particular, if $N(T)=T^{\alpha}$ with $\alpha \in (0,1)$, we get
\begin{equation}\label{g2}
\beta_{\delta_1}^+(p) \le \alpha_u^+ \le \alpha , \ p>0.
\end{equation}
(ii) Find smaller values of $N(T)$ such that $P_r(N(T), T)$ and $P_l(N(T), T)$ go to zero
according to some inverse power of $T$. This allows one to prove upper bounds
 for $\alpha_l^\pm$ and upper bounds for
$\beta^\pm(p)$ that depend on $p$ and are better than \eqref{g2}.

\medskip

Here we are mainly interested in the first problem, which deals with the sharp front of
the wavepacket.

The following results gives upper bounds on outside probabilities in terms of quantities
involving the norms of transfer matrices at complex energies.

\begin{theorem}\label{main}
Suppose $H$ is given by \eqref{oper}, where $V$ is a bounded real-valued function, and $K
\ge 4$ is such that $\sigma (H) \subseteq [-K+1,K-1]$. Then, the outside probabilities
can be bounded from above in terms of transfer matrix norms as follows:
\begin{align*}
P_r(N,T) & \lesssim \exp (-c N) + T^3 \int_{-K}^K \left( \max_{1 \le n \le
N} \left\| \Phi \left( n,E+ \tfrac{i}{T} \right) \right\|^2 \right)^{-1} dE, \\
P_l(N,T) & \lesssim \exp (-c N) + T^3 \int_{-K}^K \left( \max_{-N \le n \le -1} \left\|
\Phi \left( n,E + \tfrac{i}{T} \right) \right\|^2 \right)^{-1} dE.
\end{align*}
The implicit constants depend only on $K$.
\end{theorem}

The starting point is the Parseval formula \cite{DT,KKL},
\begin{equation}\label{g3}
a(n,T) = \frac{1}{T\pi} \int_{-\infty}^\infty \left|\langle (H - E - \tfrac{i}{T})^{-1}
\delta_1, \delta_n \rangle \right|^2 \, dE.
\end{equation}
Let us write $\e = 1/T$ and $R(z)=(H-zI)^{-1}$ for $z \in \C \setminus \R$. We assume $T
\ge 1$, so that $0<\e \le 1$. Thus, we have
\begin{equation}\label{g4}
P_r(N,T) = \frac{\e}{\pi} \int_{-\infty}^\infty M_r (N,E+i\e) \, dE,
\end{equation}
where
$$
M_r(N,z)=\sum_{n>N} |\langle R(z)\delta_1, \delta_n \rangle |^2= \| \chi_N  R(z)
\delta_1\|^2,
$$
and $\chi_N(n)=0, \ n \le N, \ \chi_N(n)=1, \ n \ge N+1$. The key problem is to bound
$M_r(N,z)$ from above. If the energy $E$ is outside the spectrum of the operator $H$,
this is a trivial problem. Assume that $E \in (-\infty, -K) \cup (K, +\infty)$, so that
$\eta={\rm dist} (E+i \e, \sigma (H)) \ge 1$. In this case the well-known Combes-Thomas
estimate yields
\begin{equation}\label{g5}
|\langle R(z)\delta_1, \delta_n \rangle | \le \frac{2}{\eta} \exp (-c \min \{ c \eta, 1
\} |n-1| )
\end{equation}
with some uniform positive $c$. Using \eqref{g5} one can easily see that
$$
\int_{E: |E| \ge K} M_r(N, E+i\e) \, dE \le C(K) \exp (-c N), \ c>0,
$$
and thus goes fast to $0$ for any $N(T)=T^{\alpha}, \ \alpha>0$. The main problem is to
estimate
$$
L_r(N,\varepsilon)=\int_{-K}^K M_r(N,E+i\e) \, dE,
$$
where $E+i\e$ may be close to the spectrum of $H$. We will link $M_r(N,z)$ to the complex
solutions to the stationary equation $Hu=zu$.

Define two operators with truncated potential:
$$
H_N^\pm \psi (n)=\psi(n-1)+\psi (n+1)+V_N^\pm(n) \psi (n),
$$
where
$$
V_N^\pm(n)=V(n), \ n \le N, \ \ V_N^\pm (n)=\pm 2K, \ n \ge N+1.
$$
Such operators already appeared (in a different context) in \cite{GKT}.
We denote by $R_N^\pm (z)$ their resolvents and define
$$
S^\pm (N,z)=\|\chi_N R_N^\pm (z) \delta_1\|^2.
$$
\begin{lemma}\label{lem1}
For any $E \in [-K,K], 0<\e<1$, we have
$$
\e^2 S^\pm (N,E+i\e) \lesssim M_r(N,E+i\e) \lesssim \e^{-2} S^\pm (N,E+i\e),
$$
where the implicit constants depend only on $K$.
\end{lemma}

\begin{proof}
By the resolvent identity, it follows that
$$
R(z)\delta_1 - R_N^\pm (z)\delta_1= R(z) (V-V_N^\pm) R_N^\pm (z) \delta_1 = R(z) (\chi_N
(V(n) \mp 2K)) R_N^\pm (z)\delta_1.
$$
Since $\|R(z)\| \le \e^{-1}, \ \e<1$, and $V$ is bounded, we get
\begin{align*}
M(N,z) & \le 2 S^\pm (N,z)+2 \|R(z)(V-V_N^\pm) R_N^\pm (z)\delta_1\|^2 \\
& \le 2 S^\pm (N,z)+2 \e^{-2} \|\chi_N (V \mp 2K) R_N^\pm (z) \delta_1\|^2 \\
& \le c_2 (K) \e^{-2} S^\pm (N,z).
\end{align*}
The other bound can be proved in a similar way.
\end{proof}

The next step is to link the quantities $S^\pm (N,z)$ with the solutions to the
stationary equation. For any complex $z$, define $u_0(n,z)$ as a solution to $Hu_0=zu_0,
\ u_0(0,z)=0, \ u_0(1,z)=1$ and $u_1(n,z)$ as a solution to $Hu_1=zu_1, \ u_1(0,z)=1, \
u_1(1,z)=0$.  An important observation is the following: since $V(n)=V_N^\pm (n), \ n \le
N$, the solutions $u_0, u_1$ for $H$ and for $H_N^\pm$ coincide for all $n \le N+1$. We
will consider $u_0,u_1$ only for such $n$, thus we use the same notation for $u_0, u_1$
in the case of $H$ and $H_N^\pm$.

For any $E \in [-K,K], \ 0<\e<1, \ z=E+i\e$, we define
$$
\lambda_{1,2}^\pm (z)= \frac{1}{2} (z \mp 2K + \sqrt{(z \mp 2K)^2 - 4}).
$$
Observe that $|z \mp 2K| \ge K \ge 4$. Thus, one of the branches of the square root
yields $|\lambda_1^\pm (z)|<1$ and the other $|\lambda_2^\pm (z)|>1$. For the values of
$z$ under consideration, we clearly have
$$
|\lambda_1^\pm (z)| \le b(K) < 1, \quad |\lambda_1^+(z)-\lambda_1^-(z)| \ge c(K) > 0
$$
with uniform constants $b(K), c(K)$. The numbers $\lambda_{1,2}^\pm (z)$
in fact are the
eigenvalues of the free equation
\begin{equation}\label{g30}
u(n-1)+u(n+1)=(z \mp 2K) u(n),
\end{equation}
whose general solution is
$$
u(n)=C_1 \lambda_1^n +C_2 \lambda_2 ^n.
$$
\begin{lemma}\label{lem2}
Let $z=E+i\e$, where $E \in [-K,K]$, $0 < \e < 1$. Then, for $n \ge N \ge 3$, we have
\begin{equation}\label{g6a}
|\langle R_N^\pm (z) \delta_1 , \delta_n \rangle| \le 2 \e^{-1} \frac{ |\lambda_1^\pm
(z)|^{n-N}}{ |\lambda_1^\pm (z) u_0(N,z) -u_0(N+1,z)| },
\end{equation}
and
\begin{equation}\label{g6b}
|\langle R_N^\pm (z) \delta_1 , \delta_n \rangle| \le \e^{-1} \frac{ |\lambda_1^\pm
(z)|^{n-N}}{ |\lambda_1^\pm (z) u_1(N,z) -u_1(N+1,z)| }.
\end{equation}
\end{lemma}

\begin{proof}
The following formula for the resolvent of any operator of the form \eqref{oper} holds
(see, e.g., \cite{KKL}):
\begin{equation}\label{g31a} \langle R(z) \delta_1, \delta_n \rangle=d(z) u_0
(n,z) +b(z) u_1 (n,z), \ n \ge 1,
\end{equation}
\begin{equation}\label{g31b}
\langle R(z) \delta_1, \delta_n \rangle=d(z) u_0 (n,z) +c(z) u_1
(n,z), \ n < 1,
\end{equation}
where
\begin{align*}
d(z) & = \frac{-m_+(z)m_-(z)}{m_+(z)+m_-(z)}, \\
b(z) & = \frac{m_-(z)}{m_+(z)+m_-(z)}, \\
c(z) & = \frac{-m_+(z)}{m_+(z)+m_-(z)},
\end{align*}
with some complex functions $m_+(z), m_-(z)$ (which depend on the potential). Since
$\|R(z) \delta_1\| \le \e^{-1}$ and
$$
d(z)=\langle R(z) \delta_1, \delta_1 \rangle , \  c(z)=\langle R(z) \delta_1, \delta_0
\rangle,
$$
we get $|d(z)| \le \e^{-1}, \ |c(z)| \le \e^{-1}$. Since $b(z)=1+c(z)$ and $\e \le 1$, we
also get $|b(z)| \le 2 \e^{-1}$. Summarizing,
\begin{equation}\label{g102}
|d(z)| \le
\e^{-1}, |c(z)| \le \e^{-1}, |b(z)| \le 2 \e^{-1}.
\end{equation}
One should stress that the bounds \eqref{g102} hold for operators of the form
\eqref{oper} with any potential, and the constants are universal.

Consider the operators $H_N^\pm$ and define $\phi^\pm=R_N^\pm (z)\delta_1$ (the vectors
$\phi^\pm$ depend on $N$, of course). Since
$$
(H_N^\pm-z) \phi^\pm=\delta_1,
$$
the function $\phi^\pm (n)=\langle \phi^\pm, \delta_n \rangle$ obeys the equation
$H_N^\pm \phi(n)=z \phi(n), n \ge 2$.  Since $V_N(n)=\pm 2K, \ n\ge N+1$, $\phi(n)$ obeys
the free equation \eqref{g30} for $n \ge N+1$. Hence,
\begin{equation}\label{g8}
(\phi^\pm
(N+k+1), \phi^\pm (N+k))^T= C_1^\pm \lambda_1^\pm (z)^k e_1+C_2^\pm \lambda_2 ^\pm(z)^k
e_2, \ k \ge 0.
\end{equation}
Here $e_{1,2}^\pm=(\lambda_{1,2}^\pm (z), 1)^T$ are two eigenvectors of the matrix
corresponding to the equation \eqref{g30}, and the constants $C_1^\pm,C_2^\pm$ are
defined by
$$
(\phi^\pm (N+1), \phi^\pm(N))^T=C_1^\pm e_1+C_2^\pm e_2.
$$
Since $|\lambda_1^\pm (z)|<1, \ |\lambda_2^\pm (z)|>1$ and $\phi^\pm  \in \ell^2(\Z)$,
the identity \eqref{g8} implies that $C_2^\pm=0$, and thus
\begin{equation}\label{g9}
(\phi^\pm (N+1), \phi^\pm (N))^T=D^\pm_N(z) (\lambda_1^\pm (z), 1)^T, \ D_N^\pm(z) \ne 0.
\end{equation}
On the other hand, \eqref{g31a} implies
\begin{equation}\label{g10a}
\phi^\pm
(N+1)=d_N^\pm (z) u_0(N+1, z) + b_N^\pm (z) u_1(N+1, z),
\end{equation}
\begin{equation}\label{g10b}
\phi^\pm (N)=d_N^\pm (z) u_0(N, z) + b_N^\pm (z) u_1(N, z),
\end{equation}
Remark that $d_N^\pm (z) \ne 0$, since it is a Borel transform of the spectral measure
corresponding to the vector $\delta_1$ \cite{KKL}. The same is true for $b_N^\pm (z),
c_N^\pm (z)$ since $m_+(z), m_-(z)$ are functions with  positive imaginary part (it also
follows from $d_N^\pm (z) \ne 0$ and the expressions of $b, c$). Using the fact that
$u_0(n+1, z) u_1(n,z)-u_0(n,z) u_1(n+1,z)=1$ for any $n$ (in particular, for $n=N$), and
\eqref{g9}--\eqref{g10b}, it is easy to calculate $D_N^\pm (z)$:
\begin{equation}\label{g104}
D_N^\pm(z) = \frac{d_N^\pm (z)}{\lambda_1^\pm (z) u_1(N,z)-u_1 (N+1,z)}=
\frac{b_N^\pm (z)}{\lambda_1^\pm (z) u_0(N,z)-u_0 (N+1,z)}.
\end{equation}
 As observed above, the solutions $u_0, u_1$ are the same for $H, H_N^\pm$ if $n \le
N+1$. It follows from \eqref{g8}, where $C_1=D^\pm$ and $C_2=0$, that
$$
\langle R_N^\pm (z) \delta_1, \delta_n \rangle =
D_N^\pm (z) \lambda_1^\pm(z)^{n-N}, \ n
\ge N.
$$
The result of the lemma follows now directly from \eqref{g102} and \eqref{g104}.
\end{proof}

\begin{lemma}\label{lem3}
For $z=E+i\e$ with $E \in [-K,K]$ and $0 < \e \le 1$, and $N \ge 3$, we have
$$
M_r(N,z) \le C(K)\e^{-4}\left( \max_{3 \le n \le N} \|\Phi(n,z)\|^2 \right) ^{-1}.
$$
\end{lemma}

\begin{proof}
The second bound of Lemma~\ref{lem1} and the bound \eqref{g6a} of Lemma~\ref{lem2} yield
\begin{align}
M_r(N,z) & \le A(K) \e^{-4} |\lambda_1^\pm (z) u_0(N,z)-u_0(N+1,z)|^{-2}
\sum_{k=0}^\infty |\lambda_1^\pm (z)|^{2k} \label{g12a} \\
& \le B(K) \e^{-4} |\lambda_1^\pm (z) u_0(N,z)-u_0(N+1,z)|^{-2} \label{g12}
\end{align}
with uniform $B(K)$, since $|\lambda_1^\pm (z)|\le b(K)<1$.

Denote $U^\pm=\lambda_1^\pm(z) u_0(N,z)-u_0(N+1,z)$. It is not hard to see that
\begin{equation}\label{g201}
|U^-|+|U^+| \ge |\lambda_1^-(z)-\lambda_1^+(z)| |u_0(N,z)|
\end{equation}
and
\begin{equation}\label{g202}
|U^-|+|U^+| \ge |\lambda_1^-(z)||U^-|+|\lambda_1^+(z)||U^+| \ge
 |\lambda_1^-(z)-\lambda_1^+(z)| |u_0(N+1,z)|.
\end{equation}
Since $|\lambda_1^-(z)-\lambda_1^+(z)|\ge c(K)>0$, \eqref{g12a}--\eqref{g202} imply
$$
M_r(N,z) \le C(K)\e^{-4} \left( |u_0(N,z)|^2+|u_0(N+1,z)|^2 \right) ^{-1}.
$$
Using \eqref{g6b}, we can prove a similar bound with $u_0$ replaced by $u_1$. Thus,
$$
M_r(N,z) \le C(K) \e^{-4} \|\Phi(N,z)\|^{-2}.
$$
Since $M_r(n,z)$ is decreasing in $n$, the asserted bound follows.
\end{proof}

\begin{proof}[Proof of Theorem~\ref{main}.]
The assertion is an immediate consequence of \eqref{g4} and Lemma~\ref{lem3} (and the
analogous results on the left half-line). Note that we can replace $[3,N]$ by $[1,N]$
since this modification only changes the $K$-dependent constant.
\end{proof}

\begin{proof}[Proof of Theorem~\ref{main1}.]
Let us choose $N(T) = \lfloor C T^\alpha \rfloor $, where $C \in (0,\infty)$ and $\alpha
\in (0,1)$ are chosen such that \eqref{assumeright} and \eqref{assumeleft} hold. Observe
that $P(N(T), T)=P(\lfloor N(T) \rfloor , T)$. Then Theorem~\ref{main} shows that
$P_r(N(T),T)$ and $P_l(N(T),T)$ go to $0$ faster than any inverse power of $T$. By
definition of $S^+(\alpha)$ and $\alpha_u^+$ (cf.~\eqref{spludef} and \eqref{audef}), it
follows that $\alpha_u^+ \le \alpha$, which is \eqref{apubound}. Finally,
\eqref{bppbound} follows from \eqref{betaalphabound}.
\end{proof}

Let us now show that the transport exponents $\alpha_u^\pm$ can be expressed in terms of
the integrals
$$
I(N, T)=\int_{-K}^K \left( \left\|\Phi \left( N, E+\tfrac{i}{T} \right) \right\|^{-2} +
\left\| \Phi \left( -N, E+\tfrac{i}{T} \right) \right\|^{-2} \right) \, dE.
$$
It follows from Theorem~\ref{main} that
\begin{equation}\label{h10}
P(N,T)  \lesssim \exp (-cN)+T^3 I(N,T).
\end{equation}
Define
\begin{align*}
W^-(\alpha) & = - \liminf_{T \to \infty} \frac{\log I(\lfloor T^\alpha -2 \rfloor, T)
}{\log T}, \\
W^+(\alpha) & = - \limsup_{T \to \infty} \frac{\log I(\lfloor T^\alpha -2 \rfloor, T)
}{\log T}, \\
\gamma^\pm  & = \sup \{ \alpha \ge 0  :  W^\pm (\alpha) < \infty \}.
\end{align*}

\begin{theorem}\label{I}
We have that
$$
\alpha_u^\pm  =  \gamma^\pm.
$$
\end{theorem}

\begin{proof}
We first prove a lower bound for $a(n+1,T)+a(n,T)$. Let $\phi (n)=\langle R(z) \delta_1,
\delta_n \rangle$, where $z=E+i/T$. The identities \eqref{g31a}, \eqref{g31b} imply
\begin{align*}
(\phi(n+1), \phi(n))^T & = \Phi (n, z) (d(z), b(z))^T, \ n \ge 1, \\
(\phi(n+1), \phi(n))^T & = \Phi (n, z) (d(z), c(z))^T, \ n < 1.
\end{align*}
Since $\mathrm{det} \Phi (n,z)=1$, we obtain (see, e.g., \cite{DT} for details)
\begin{equation}\label{h14}
|\phi(n+1)|^2+|\phi(n)|^2 \ge \|\Phi(n,z)\|^{-2} |d(z)|^2
\end{equation}
for any $n$. Since $d(z)=\langle R(z) \delta_1, \delta_1 \rangle$ is the Borel transform
of a compactly supported measure, it is easy to see that $|d(E+\tfrac{i}{T})|\ge
\mathrm{Im} \, d(E+\tfrac{i}{T}) \ge C/T$ with a constant that is uniform in $E \in
[-K,K],T \ge 1$. Therefore, \eqref{g3} and \eqref{h14} yield
$$
a(n+1, T)+a(n,T) \ge \frac{C}{T^3} \int_{-K}^K \left\| \Phi \left( n,E+\tfrac{i}{T}
\right) \right\|^{-2} \, dE,
$$
and
\begin{equation}\label{h11}
P(N,T) \ge \frac{C}{T^3} \int_{-K}^K  \left( \left\| \Phi \left( N+1, E+\tfrac{i}{T}
\right) \right\|^{-2} + \left\| \Phi \left(-N-2, E+\tfrac{i}{T} \right) \right\|^{-2}
\right) \, dE.
\end{equation}
Since the potential $V$ is bounded, and $E \in [-K, K], \ T \ge 1$, we have that
\begin{align*}
\|\Phi(N+1, E+\tfrac{i}{T})\| & \asymp \|\Phi(N,E+\tfrac{i}{T})\| , \\ \|\Phi(-N-2,
E+\tfrac{i}{T})\| & \asymp \|\Phi(-N, E+\tfrac{i}{T})\|,
\end{align*}
with constants depending only on $K$. Therefore, (\ref{h11}) implies
\begin{equation}\label{h12}
P(N,T) \ge \frac{C}{T^3} I(N,T).
\end{equation}
Let us show that $\alpha_u^+=\gamma^+$. Assume first that $\gamma^+<\alpha_u^+$. Then
there exists $\alpha$ such that $\gamma^+ < \alpha < \alpha_u^+$. The definition of
$\gamma^+$ implies $W^+(\alpha) = \infty$ and thus
\begin{equation}\label{h13}
I(\lfloor T^\alpha -2 \rfloor, T) \le \frac{C_M}{T^M}.
\end{equation}
for any $M>0$ uniformly in $T$. It follows from \eqref{h10} that $P(T^\alpha-2,
T)=P(\lfloor T^\alpha -2 \rfloor, T)$ also obeys a bound like \eqref{h13}, and thus
$S^+(\alpha) = \infty$. However, this is impossible since $\alpha<\alpha_u^+$. Therefore,
$\gamma^+ \ge \alpha_u^+$. Similarly, $\gamma^+ \le \alpha_u^+$ follows from the bound
\eqref{h12}. The identity $\alpha_u^-=\gamma^-$ can be proven in the same way (with
inequalities like \eqref{h13} valid for some sequence of times $T_k \to \infty$).
\end{proof}

\begin{proof}[Proof of Theorem~\ref{stab}.]
It follows directly from (\ref{h16}) and the definition of $W^\pm (\alpha)$ that
$|W_1^\pm (\alpha)-W_2^\pm (\alpha)| \le 2A$ for any $\alpha$, and hence
$\gamma_1^\pm=\gamma_2^\pm$. The result now follows from Theorem~\ref{I}. (Since
$\alpha_u^\pm \le 1$, it is sufficient that \eqref{h16} holds for $|n| \le 1/\e$.)
\end{proof}

\section{The Fibonacci Operator}

In this section we prove upper bounds for the transport exponents $\beta^+(p)$ in the
case of the Fibonacci potential given by \eqref{fibpot} for $\lambda$ sufficiently large.

In order to apply Theorem~\ref{main}, we need to find lower bounds for the norms of
transfer matrices at complex energies. Rather than norms, we will study traces of
transfer matrices. This will be sufficient because lower bounds for traces give rise to
lower bounds for norms in a trivial way. Moreover, the study of traces is natural in the
case of the Fibonacci model since it displays a well-known hierarchical structure on the
level of transfer matrices that induces recursion relations for the traces of transfer
matrices over intervals whose length is given by a Fibonacci number. This in turn is most
conveniently described by a dynamical system, the so-called trace map. We remark that in
the existing proofs of lower bounds for transport exponents, it was required to find
upper bounds for transfer matrix norms. Interestingly enough, it was sufficient also in
that case to study traces. This less trivial transition from traces to norms can be
achieved in the situation at hand using an observation of Iochum and Testard \cite{it};
see also \cite{irt}.

While the trace map has been studied by many authors (e.g.,
\cite{bist,it,KKL,kkt,oprss,r,su}), it has been analyzed so far only for real energies.
Thus, the trace map dynamical system is usually considered as a map from $\R^3$ to
$\R^3$. Theorem~\ref{main} suggests studying complex energies. Thus, we are led to study
the trace map as a complex dynamical system, that is, as a map from $\C^3$ to $\C^3$. The
initial element of $\C^3$ will be an energy-dependent vector. We investigate the stable
set, that is, the set of complex energies such that the associated vector in $\C^3$ has a
bounded trace map orbit. This set of energies is, in fact, a subset of $\R$, but it is
important to study the canonical approximants to this set as subsets of $\C$. Their
complements are escaping regions in the sense that once an orbit enters such a set, it
escapes to infinity at a super-exponential rate. Since all non-real energies give rise to
escaping trace map orbits, we can study in this way the maximum number of iterates it
takes, for a given imaginary part $\varepsilon$ of the energy $z$, to enter an escaping
region. From this point on, we can control the increase of the trace and this will prove
sufficient for us to obtain the lower bounds we need.

Our goal is to derive Theorem~\ref{main2} from Theorem~\ref{main1}. Since $V(-n) =
V(n-1)$ for $n \ge 2$ (see \cite{su}), we can restrict our attention to one half line.
Thus, we will give details only for \eqref{assumeright} with suitable $\alpha$. The proof
of \eqref{assumeleft} is completely analogous.

For $z \in \C$, define the matrices $M_k(z)$ by
\begin{equation}\label{fibtransfer}
\Phi(F_k,z) = M_k(z), \quad k \ge 1,
\end{equation}
 where $F_k$ is the $k$-th Fibonacci number, that is, $F_0 = F_1 =
1$ and $F_{k+1} = F_k + F_{k-1}$ for $k \ge 1$. It is well-known that
$$
M_{k+1}(z) = M_{k-1}(z) M_k(z), \quad k \ge 2.
$$
For the variables $x_k(z) = (\mathrm{Tr} \ M_k(z))/2$, $k \ge 1$, we have the recursion
\begin{equation}\label{tmrecursion}
x_{k+1}(z) = 2 x_{k}(z) x_{k-1}(z) - x_{k-2}(z)
\end{equation}
and the invariant
\begin{equation}\label{inv}
x_{k+1}(z)^2 + x_k(z)^2 + x_{k-1}(z)^2 - 2x_{k+1}(z) x_k(z) x_{k-1}(z) - 1 \equiv
\frac{\lambda^2}{4}.
\end{equation}
Letting $x_{-1}(z) = 1$ and $x_0(z) = z/2$, the recursion \eqref{tmrecursion} holds for
all $k \ge 0$. See, for example, \cite{kkt,oprss,su}, for these results.

For $\delta \ge 0$, consider the sets
$$
\sigma_k^\delta = \{ z \in \C : |x_k(z)| \le 1 + \delta \}.
$$
For $S \subseteq \C$, we let $S^\R = S \cap \R$. The set $\sigma_k^0$ is the spectrum of
the $k$-th periodic approximant to the Fibonacci operator. In other words, $x_k$ is the
discriminant of this periodic operator and hence the set $(\sigma_k^0)^\R$ consist of
$F_k$ closed bands that, in general, have disjoint interiors. However, Raymond has shown
that in these particular cases, even the closed bands are mutually disjoint \cite{r}.
Thus, there are exactly $F_k$ disjoint bands making up the set $(\sigma_k^0)^\R$, and
each one of them contains exactly one zero of $x_k$ and each one of the $F_k - 1$ bounded
gaps contains exactly one critical point of $x_k$.

We have
$$
\sigma_k^\delta \cup \sigma_{k-1}^\delta \supseteq \sigma_{k+1}^\delta \cup
\sigma_k^\delta \to \sigma,
$$
the latter set being the spectrum of the Fibonacci Hamiltonian. Outside of these sets,
the traces grow super-exponentially. More precisely, a modification of S\"ut\H{o}'s proof
shows the following:

\begin{lemma}\label{suto}
A necessary and sufficient condition that $x_k(z)$ be unbounded is that
\begin{equation}\label{critN}
|x_{N-1}(z)| \le 1 + \delta, \quad |x_{N}(z)| > 1 + \delta, \quad |x_{N+1}(z)| > 1 +
\delta
\end{equation}
for some $N \ge 0$. This $N$ is unique. Moreover, in this case we have
\begin{equation}\label{supmult}
|x_{n+2}(z)| > |x_{n+1}(z) x_n(z)| \text{ for } n \ge N
\end{equation}
and
\begin{equation}\label{expgrowth}
|x_{N+n}(z)| \ge (1 + \delta)^{F_n} \text{ for } n \ge 0.
\end{equation}
\end{lemma}

\begin{proof}
Suppose that \eqref{critN} holds true for some $N \ge 0$. Then
\begin{align*}
|x_{N+2}(z)| & \ge |x_{N+1}(z) x_{N}(z)| + \left( |x_{N+1}(z) x_{N}(z)| - |x_{N-1}(z)|
\right)\\
& \ge |x_{N+1}(z) x_{N}(z)| + \delta(1 + \delta)
\end{align*}
This also implies $|x_{N+2}(z) x_{N+1}(z)| > |x_{N}(z)|$. Thus, \eqref{supmult} follows
by induction. This in turn shows both \eqref{expgrowth} and that there is at most one $N
\ge 0$ with \eqref{critN}.

Now suppose that \eqref{critN} holds for no value of $N \ge 0$. Since $x_{-1}(z) = 1$,
this implies that for every $n$ with $|x_n(z)| > 1 + \delta$, we must have both
$|x_{n-1}(z)| \le 1 + \delta$ and $|x_{n+1}(z)| \le 1 + \delta$. The invariant
\eqref{inv} therefore shows that the sequence $x_n(z)$ must be bounded.
\end{proof}

Moreover, assuming $\lambda > \lambda_0(\delta)$, where
\begin{equation}\label{lambda0}
\lambda_0(\delta) = [12(1 + \delta)^2 + 8(1 + \delta)^3 + 4]^{1/2},
\end{equation}
the invariant \eqref{inv} implies
\begin{equation}\label{disj}
\sigma_k^\delta \cap \sigma_{k+1}^\delta \cap \sigma_{k+2}^\delta = \emptyset.
\end{equation}

\begin{lemma}\label{deltareal}
Fix $\delta \ge 0$ and $\lambda > \lambda_0(\delta)$. Then the set $(\sigma_k^\delta)^\R$
consists of $F_k$ closed bands that are mutually disjoint. Each one of them contains
exactly one zero of $x_k$ and each one of the $F_k - 1$ bounded gaps contains exactly one
critical point of $x_k$.
\end{lemma}

\begin{proof}
For $\delta = 0$, these statements are known and were recalled above. If $\delta > 0$,
let $t$ run from $0$ to $\delta$. The bands of $(\sigma_k^t)^\R$ fatten up and what needs
to be avoided is that two consecutive bands touch. By the assumption $\lambda >
\lambda_0(\delta)$ this is impossible, however, as can be seen from \eqref{disj} and the
structure of the sets $(\sigma_k^\delta)^\R$ induced by this property (compare Figure~1
and the discussion of the sets $(\sigma_k^0)^\R$ in \cite[Section~5]{KKL} and
\cite[Section~6]{r}).
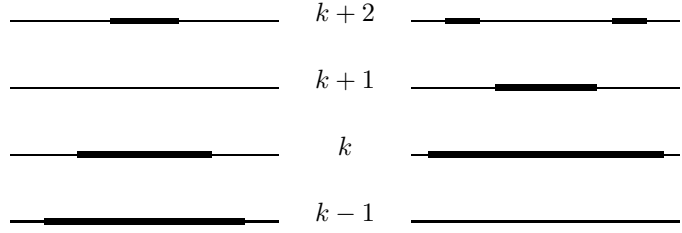
\begin{figure}\label{Fig1}
\begin{center}
\setlength{\unitlength}{0.175in}
\begin{picture}(20,7)(0,0)
\put(0,0){\line(1,0){8}} \put(10,0){\hbox to 0mm{\hss$k-1$\hss}}
\put(12,0){\line(1,0){8}} \put(0,2){\line(1,0){8}} \put(10,2){\hbox to 0mm{\hss$k$\hss}}
\put(12,2){\line(1,0){8}} \put(0,4){\line(1,0){8}} \put(10,4){\hbox to
0mm{\hss$k+1$\hss}} \put(12,4){\line(1,0){8}} \put(0,6){\line(1,0){8}} \put(10,6){\hbox
to 0mm{\hss$k+2$\hss}} \put(12,6){\line(1,0){8}}
\linethickness{2pt} \put(1,0){\line(1,0){6}} \put(2,2){\line(1,0){4}}
\put(3,6){\line(1,0){2}}
\put(12.5,2){\line(1,0){7}} \put(14.5,4){\line(1,0){3}} \put(13,6){\line(1,0){1}}
\put(18,6){\line(1,0){1}}
\end{picture}
\end{center}
\caption{The structure of the sets $(\sigma_k^\delta)^\R$ when \eqref{disj} holds.}
\end{figure}
\end{proof}

\begin{lemma}
Fix $\delta \ge 0$ and $\lambda > \lambda_0(\delta)$. Then the set $\sigma_k^\delta$ has
exactly $F_k$ connected components. Each of them is a topological disk that is symmetric
about the real axis.
\end{lemma}

\begin{proof}
Since $x_k$ has degree $F_k$, the set $\sigma_k^\delta$ has at most $F_k$ connected
components. Since the coefficients of $x_k$ are real, it follows also that each such
component must be symmetric about the real axis. On the other hand, Lemma~\ref{deltareal}
shows that the set $(\sigma_k^\delta)^\R$ has exactly $F_k$ connected components. Thus,
the maximum modulus principle show that each component of $\sigma_k^\delta$ is a
topological disk and there are exactly $F_k$ of them.
\end{proof}

The final ingredients we need are the following bounds on $|x_k'(z)|$ for $z \in
(\sigma_k^0)^\R$ (cf.~\cite[Proposition~5.2]{KKL} and \cite[Equation~(57)]{DT}):

\begin{prop}\label{kklprop}
If $\lambda > 8$, then
\begin{equation}\label{xprimebound}
|x_k'(z)| \ge \xi(\lambda)^{k/2}
\end{equation}
for all $k \ge 3$ and $z \in (\sigma_k^0)^\R$. Here,
$$
\xi(\lambda) = \frac{\lambda - 4 + \sqrt{(\lambda - 4)^2 - 12}}{2}
$$
and hence $\xi(\lambda) = \lambda + O( 1 )$ as $\lambda \to \infty$.
\end{prop}

\begin{prop}\label{dtprop}
If $\lambda > 4$, then
\begin{equation}\label{xprimebound2}
|x_k'(z)| \le C (2 \lambda + 22)^k
\end{equation}
for all $k \ge 1$ and $z \in (\sigma_k^0)^\R$.
\end{prop}

Now we can prove our main result on the structure of the set $\sigma_k^\delta$. Denote
$B(z,r) = \{ w \in \C : | w - z | < r \}$.

\begin{prop}
Fix $k \ge 3$, $\delta > 0$, and $\lambda > \lambda_0(2\delta)$. Then, there are
constants $c_\delta,d_\delta > 0$ such that
\begin{equation}\label{unionballs}
\bigcup_{j = 1}^{F_k} B \left( z_k^{(j)}, r_k \right) \subseteq \sigma_k^\delta \subseteq
\bigcup_{j = 1}^{F_k} B \left( z_k^{(j)}, R_k \right),
\end{equation}
where $\{z_k^{(j)}\}_{1 \le j \le F_k}$ are the zeros of $x_k$, $r_k = c_\delta (2\lambda
+ 22)^{-k}$, and $R_k = d_\delta \xi(\lambda)^{-k/2}$.
\end{prop}

\begin{proof}
Consider a connected component $C_j$ of $\sigma_k^{2\delta}$. By the assumption  $\lambda
> \lambda_0(2\delta)$, $C_j$ contains exactly one of the $F_k$ real zeros, $z_k^{(j)}$.
Moreover, it contains exactly one connected component of $\sigma_k^\delta$, which we
denote by $\tilde{C}_j$. Our goal is to show that with the radii $r_k,R_k$ described in
the proposition,
\begin{equation}\label{cjgoal}
B \left( z_k^{(j)}, r_k \right) \subseteq \tilde{C}_j \subseteq B \left( z_k^{(j)}, R_k
\right).
\end{equation}
Clearly, \eqref{unionballs} follows from \eqref{cjgoal}.

Since $C_j$ contains exactly one zero of $x_k$, it follows from the Maximum Modulus
Theorem and Rouch\'e's Theorem (e.g., \cite[Theorems~6.13 and 10.10]{bn}) that
$$
x_k : \mathrm{int}(C_j) \to B(0,1 + 2\delta)
$$
is univalent, and hence
$$
x_k^{-1} : B(0,1 + 2\delta) \to \mathrm{int}(C_j)
$$
is well-defined and univalent as well. Consequently, the following mapping is a Schlicht
function:
$$
F : B(0,1) \to \C, \quad F(z) = \frac{x_k^{-1} ((1 + 2\delta)z) - z_k^{(j)}}{(1 +
2\delta) [(x_k^{-1})'(0)]}.
$$
That is, $F$ is a univalent function on $B(0,1)$ with $F(0) = 0$ and $F'(0) = 1$.

The Koebe Distortion Theorem (see \cite[Theorem~7.9]{c}) implies that
\begin{equation}\label{koebe}
\frac{|z|}{(1 + |z|)^2} \le |F(z)| \le \frac{|z|}{(1 - |z|)^2} \text{ for } |z| \le 1.
\end{equation}
Evaluate the bound \eqref{koebe} on the circle $|z| = \frac{1 + \delta}{1 + 2 \delta}$.
For such $z$, we obtain
$$
\frac{(1 + \delta)(1 + 2\delta)}{(2 + 3 \delta)^2} \le |F(z)| \le \frac{(1 + \delta)(1 +
2 \delta)}{\delta^2}.
$$
By definition of $F$ this means that
$$
| x_k^{-1} ((1 + 2\delta)z) - z_k^{(j)} | \le \frac{(1 + \delta)(1 + 2 \delta)}{\delta^2}
(1 + 2\delta) |(x_k^{-1})'(0)|
$$
and
$$
| x_k^{-1} ((1 + 2\delta)z) - z_k^{(j)} | \ge \frac{(1 + \delta)(1 + 2 \delta)}{(2 +
3\delta)^2} (1 + 2\delta) |(x_k^{-1})'(0)|
$$
for all $z$ with $|z| = \frac{1 + \delta}{1 + 2 \delta}$. In other words, if $|z| = 1 +
\delta$, then
\begin{equation}\label{variation}
| x_k^{-1} (z) - z_k^{(j)} | \le \frac{(1 + \delta)(1 + 2 \delta)^2}{\delta^2}
|(x_k^{-1})'(0)|
\end{equation}
and
\begin{equation}\label{variation2}
| x_k^{-1} (z) - z_k^{(j)} | \ge \frac{(1 + \delta)(1 + 2 \delta)^2}{(2 + 3\delta)^2}
|(x_k^{-1})'(0)|
\end{equation}
Since $|(x_k^{-1})'(0)| = |x_k'(z_k^{(j)})|^{-1}$, we obtain from
Proposition~\ref{kklprop} and \eqref{variation}
\begin{equation}\label{distance1}
| x_k^{-1} (z) - z_k^{(j)} | < \left(\frac{(1 + \delta)(1 + 2 \delta)}{\delta} \right)^2
\xi(\lambda)^{-k/2}
\end{equation}
for all $z$ of magnitude $1 + \delta$. Similarly, Proposition~\ref{dtprop} and
\eqref{variation2} give
\begin{equation}\label{distance2}
| x_k^{-1} (z) - z_k^{(j)} | \ge \frac{(1 + \delta)(1 + 2 \delta)^2}{(2 + 3\delta)^2}
\frac{1}{C} (2 \lambda + 22)^{-k}
\end{equation}
for these values of $z$. Note that as $z$ runs through the circle of radius $1 + \delta$
around zero, the point $x_k^{-1} (z)$ runs through the entire boundary of $\tilde{C}_j$.
Thus, \eqref{cjgoal} follows from \eqref{distance1} and \eqref{distance2}.
\end{proof}

\begin{proof}[Proof of Theorem~\ref{main2}.]
By the assumption $\lambda > 8$, it is possible to choose $\delta > 0$ such that $\lambda
> \lambda_0(2\delta)$; compare \eqref{lambda0}. Since the $z_k^{(j)}$ are real, the last
proposition gives in particular that
$$
\sigma_k^\delta \subseteq \{ z \in \C : |\mathrm{Im} \ z| < d_\delta \xi(\lambda)^{-k/2}
\} \subseteq \{ z \in \C : |\mathrm{Im} \ z| < d_\delta F_k^{-\gamma(\lambda)} \}.
$$
for a suitable $\gamma(\lambda)$. This implies that
\begin{equation}\label{imwidth}
\sigma_k^\delta \cup \sigma_{k+1}^\delta \subseteq \{ z \in \C : |\mathrm{Im} \ z| <
d_\delta F_k^{-\gamma(\lambda)} \}.
\end{equation}

If $\eta = (\sqrt{5}+1)/2$ (so that $F_k$ behaves like $\eta^k$ for $k$ large enough),
then the constant $\gamma(\lambda)$ may be chosen according to
$$
\gamma(\lambda) = \frac{\log \xi (\lambda) }{2 (1 + \nu) \log \eta}
$$
for some $\nu > 0$. Since we are interested in large $T$ behavior, we can work with any
positive $\nu$ and then have the inclusion above for $k \ge k_0(\nu)$.

In other words, for each $\varepsilon = \mathrm{Im} \ z > 0$, one obtains lower bounds on
$|x_n(E+i\varepsilon)|$ which are uniform for $E \in [-K,K] \subseteq \R$. Namely, given
$\varepsilon > 0$, choose $k$ minimal with the property $d_\delta F_k^{-\gamma(\lambda)}
< \varepsilon$. By \eqref{imwidth}, we infer that $|x_k(E+i\varepsilon)| > 1 + \delta$
and $|x_{k+1}(E+i\varepsilon)| > 1 + \delta$. Since $|x_{-1}(E+i\varepsilon)| = 1 \le 1 +
\delta$, we must have the situation of Lemma~\ref{suto} for some $N \le k$. In
particular, for $n > k$, \eqref{expgrowth} shows that
$$
|x_n(E+i\varepsilon)| \ge (1 + \delta)^{F_{n-k}}.
$$

This motivates the following definitions. Fix some small $\delta > 0$. For $T > 1$,
denote by $k(T)$ the unique integer with
$$
\frac{F_{k(T) - 1}^{\gamma(\lambda)}}{d_\delta} \le T <
\frac{F_{k(T)}^{\gamma(\lambda)}}{d_\delta}
$$
and let
$$
N(T) = F_{k(T) + \lfloor \sqrt{k(T)} \rfloor}.
$$
Thus, for every $\tilde \nu > 0$, there is a constant $C_{\tilde \nu} > 0$ such that
\begin{equation}\label{ntfib}
N(T) \le C_{\tilde \nu} T^\frac{1}{\gamma(\lambda)} T^{\tilde \nu}.
\end{equation}
It follows from Theorem~\ref{main} and the argument above that
\begin{align*}
P_r(N(T),T) & \lesssim \exp (-c N(T)) + T^3 \int_{-K}^K \left( \max_{3 \le n \le N(T)}
\left\| \Phi \left( n,0; E+\tfrac{i}{T} \right) \right\|^2 \right) ^{-1} dE \\
& \lesssim \exp (-c N(T)) + T^3 (1 + \delta)^{-2 F_{\lfloor \sqrt{k(T)} \rfloor}}.
\end{align*}
From this bound, we see that $P_r(N(T),T)$ goes to zero faster than any inverse power. We
note again that we can obtain the same result on the left half-line due to the symmetry
of the potential. Therefore, by \eqref{ntfib}, we can apply Theorem~\ref{main1} with
$$
\alpha = \frac{1}{\gamma(\lambda)} + \tilde \nu = \frac{2 (1 + \nu) \log \eta}{\log \xi
(\lambda)} + \tilde \nu,
$$
and hence \eqref{fibconcl} and \eqref{fibconcl2} follow from \eqref{apubound} and
\eqref{bppbound}, respectively, since we can take $\nu$ and $\tilde \nu$ arbitrarily
small.
\end{proof}

\section{The Almost Mathieu Operator}

Suppose we are given an operator family $\{ H_{\theta,\omega} \}$ satisfying the
assumptions of Theorem~\ref{amothm}.

Let $K = \|f\|_\infty + 3$. Then,
\begin{equation}\label{specinint}
\sigma(H_{\theta,\omega}) \subseteq [-K+1,K-1] \quad \text{ for all } \omega \in [0,1).
\end{equation}

We will show the following

\begin{prop}\label{amoprop}
Assume that $H_{\theta,\omega}$ obeys the assumptions of Theorem~\ref{amothm}. Denote
\begin{align*}
I_r(N,T) & = \int_{-K}^K \left( \max_{1 \le n \le N} \left\| \Phi \left( n,E+
\tfrac{i}{T}, \theta, \omega \right) \right\|^2 \right)^{-1} dE, \\
I_l(N,T) & = \int_{-K}^K \left( \max_{1 \le -n \le N} \left\| \Phi \left(
n,E+\tfrac{i}{T}, \theta, \omega \right) \right\|^2 \right)^{-1} dE.
\end{align*}
{\rm (a)} For every $\alpha > 0$, there exists a sequence $T_k \to \infty$ such that
\begin{equation}\label{amopropright}
I_r(T_k^\alpha, T_k)= O(T_k^{-m})
\end{equation}
and
\begin{equation}\label{amopropleft}
I_l(T_k^\alpha, T_k)= O(T_k^{-m})
\end{equation}
for every $m \ge 1$.\\
{\rm (b)} Assume moreover that $\theta$ obeys \eqref{weakbrjuno}. Then for any
$\alpha>0$,
\begin{equation}\label{dio}
I_r(T^\alpha, T)=O(T^{-m}), \ I_l(T^{\alpha}, T)=O(T^{-m})
\end{equation}
for every $m \ge 1$.
\end{prop}

\noindent\textit{Remarks.} (a) The $\alpha$-dependent sequence $\{T_k\}$ will be related
to the denominators $q_k$ of the continued fraction approximants $p_k/q_k$ of $\theta$
(cf.~\cite{khin}).
\\
(b) Our proof of Proposition~\ref{amoprop} will have some elements in common with the
arguments given by Jitomirskaya and Last in \cite{JL2} in their proof of
zero-dimensionality of spectral measures.

\begin{proof}[Proof of Theorem~\ref{amothm}.]
The assertion is an immediate consequence of Proposition~\ref{amoprop} together with
Theorem~\ref{main1} and the first remark after that theorem.
\end{proof}

\begin{proof}[Proof of Proposition~\ref{amoprop}.]
Denote
$$
\mathcal{R} = \{ z \in \C : -K \le \Re z \le K , \; 0 \le \Im z \le 1 \}.
$$
It is easy to see that there exists a constant $\Gamma'$ such that
\begin{equation}\label{gammaupper}
\|\Phi(n,z,\theta,\omega) \| \le e^{|n|\Gamma'} \text{ for all } \omega \in [0,1), \, n
\in \Z , \, z \in \mathcal{R}.
\end{equation}
Combining this with the assumption of Theorem~\ref{amothm}, we obtain
\begin{equation}\label{gammatwosided}
0 < \Gamma \le \gamma_\theta(z) \le \Gamma' < \infty \text{ for all } z \in \mathcal{R}.
\end{equation}
For $n \ge 1$, let
$$
A_n = \left\{ \omega \in [0,1) : \|\Phi(n,z,\theta,\omega) \| > \exp \left( \frac{n
\gamma_\theta(z)}{2}\right) \right\}.
$$
It follows from \eqref{lyapex} and \eqref{gammaupper} that
\begin{align*}
n \gamma_\theta(z) & \le \int_0^1 \log \|\Phi(n,z,\theta,\omega) \| \, d\omega \\
& = \int_{A_n} \log \|\Phi(n,z,\theta,\omega) \| \, d\omega + \int_{[0,1) \setminus A_n}
\log \|\Phi(n,z,\theta,\omega) \| \, d\omega \\
& \le |A_n| n \Gamma' + (1 - |A_n|) n \frac{\gamma_\theta(z)}{2},
\end{align*}
where $| \cdot |$ denotes Lebesgue measure.

Therefore,
$$
|A_n| \ge \frac{\gamma_\theta(z)}{2\Gamma' - \gamma_\theta(z)} \ge \frac{\Gamma}{2\Gamma'
- \Gamma} \equiv c.
$$
Let $d$ be the degree of the trigonometric polynomial $f$.\footnote{Note that we have $d
\ge 1$ since the assumptions of the theorem preclude the case $d = 0$.} Then each entry
of $\Phi(n,z,\theta,\omega)^* \Phi(n,z,\theta,\omega)$ is a trigonometric polynomial of
degree at most $2nd$, and the set $A_n$ consists of no more than $4nd$ intervals.
Therefore, there exists an interval $I_n \subseteq A_n$ with\footnote{If one considers
the Hilbert-Schmidt norm $\|\Phi\|_2^2 = \mathrm{Tr} (\Phi^* \Phi)$, which is of course
equivalent to the operator norm, this claim is obvious.}
$$
|I_n| \ge \frac{c}{4nd}.
$$
Let
$$
n_k = \left\lfloor \frac{c q_k}{4d} \right\rfloor + 1.
$$
By a standard lemma in continued fraction approximation (see, e.g., \cite[Lemma~9]{JL2}
for a formulation suitable to our purpose), for every $\omega$, there exists a $j \in
\{0,1,\ldots,q_k + q_{k-1} -1\}$ such that $j\theta + \omega \mod 1$ belongs to
$I_{n_k}$, and hence to $A_{n_k}$. That is, $\|\Phi(n_k,z,\theta,j\theta + \omega) \| >
\exp \left( \frac{n_k \gamma_\theta(z)}{2}\right)$. Since
$$
\Phi(j+n_k,z,\theta,\omega) = \Phi(n_k,z,\theta,j\theta + \omega) \Phi(j,z,\theta,\omega)
$$
and each $\Phi$ is unimodular, we see that either
$\|\Phi(j+n_k, z, \theta, \omega)\|$ or
$\| \Phi(j,z,\theta,\omega) \|$ is greater than $\exp \left( \frac{n_k
\gamma_\theta(z)}{4}\right)$.

Let
$$
j_k = \min \left\{ j \in \{ 0,\ldots,q_k + q_{k-1} - 1 + n_k \} : \|
\Phi(j,z,\theta,\omega) \| > \exp \left( \frac{n_k \gamma_\theta(z)}{4}\right) \right\}.
$$
By definition of $j_k$, \eqref{gammaupper}, and \eqref{gammatwosided}, we have $\frac{n_k
\Gamma}{4\Gamma'} \le j_k$ and hence there are $z$-independent constants $C_1,C_2$ such
that
\begin{equation}\label{jkbounds}
C_1 q_k \le j_k \le C_2 q_k .
\end{equation}
Moreover, there is a $z$-independent constant $C_3$ such that
\begin{equation}\label{amotmlower}
\|\Phi(j_k,z,\theta,\omega) \| > \exp \left( C_3 q_k \right)
\end{equation}
Thus, the definition of $I_r(N,T)$ and \eqref{amotmlower} imply
\begin{equation}\label{upI}
I_r(C_2 q_k, T) \le 2K \exp (-2C_3 q_k)
\end{equation}
for any $T \ge 1, k \in \N$. To prove part (a) of the proposition, consider, for a given
$\alpha>0$, the sequence $T_k = [C_2 q_k]^{1/\alpha}$. The bound \eqref{amopropright}
follows from \eqref{upI}. Since
\begin{equation}\label{symmetry}
\Phi(n,z,\theta,\omega) = \Phi(-n,z,-\theta,\theta + \omega)
\end{equation}
and the $q_k$'s are the same for $\theta$ and $-\theta$, we see that we can prove
\eqref{amopropleft} with the same sequence $\{T_k\}$.

Let us now prove part (b) of the proposition. Choose some $\alpha>0$. For any given $T
\ge 1$, one can find some $k=k(T)$ such that
\begin{equation}\label{dio2}
(C_2q_k)^{1/\alpha} \le T \le (C_2q_{k+1})^{1/\alpha}.
\end{equation}
Let
\begin{equation}\label{dio3}
\gamma_k=\frac{\log q_{k+1}}{q_k}.
\end{equation}
It follows from the definition of $I_r$, \eqref{upI} and \eqref{dio2}--\eqref{dio3} that
$$
I_r(T^\alpha, T) \le I_r(C_2 q_k, T) \le 2K \exp (-2 C_3 q_k) = 2K q_{k+1}^{-2
C_3/\gamma_k}\le C T^{-2 \alpha C_3/\gamma_k}.
$$
By assumption we have $\lim_{k \to \infty} \gamma_k=0$. Since $k(T) \to \infty$ as $T \to
\infty$, we get $I_r(T^\alpha, T)=O(T^{-m})$ for every $m \ge 1$. Due to the observation
\eqref{symmetry}, the same is true for $I_l(T^\alpha, T)$. This gives \eqref{dio} and
concludes the proof.
\end{proof}

\section{Concluding Remarks}

We introduced a general upper bound on outside probabilities in terms of transfer matrix
norms at complex energies. We remark that a completely analogous proof shows the same
result for operators of the form \eqref{oper} in $\ell^2(\Z_+)$ with any self-adjoint
boundary condition at zero (e.g., Dirichlet). This upper bound for outside probabilities
can be used to bound the transport exponents from above in certain situations.

While we carried out the analysis for the initial state $\delta_1$, one can prove
dynamical upper bounds for more general initial states. Namely, one can take $\psi = f(H)
\delta_1$, where $f$ is some smooth compactly supported function, and apply ideas from
\cite{T}. In this way one can study dynamics locally in the energy. For example, if $H$
has some absolutely continuous spectrum, but the spectrum is purely singular on $[a,b]$,
one can consider functions $f$ that are supported on $[a,b]$, so that $\psi$ belongs to
the singular subspace of $H$.

The general result was used to prove quantum dynamical upper bounds for the Fibonacci
operator in the large-coupling regime. This constitutes the first example for which a
non-trivial upper bound on $\beta^+_{\delta_1}(p)$, $0 < p < \infty$, can be established
outside the realm of dynamically localized systems. In particular, we obtain the first
rigorous anomalous transport result. The asymptotic dependence on the coupling parameter
follows a law that was numerically predicted by Abe and Hiramoto in 1987.

Combining the ideas of the present paper with those from \cite{d2}, it is possible to
perform a similar quantum dynamical analysis for general Sturmian potentials given by
\eqref{sturm}.

Moreover, it is possible to improve the known dynamical lower bounds for the Fibonacci
Hamiltonian by employing the first inclusion in \eqref{unionballs} together with the main
result of \cite{DT}.

We also showed that all lower transport exponents and, under a weak assumption on the
frequency $\theta$, all upper transport exponents vanish for quasi-periodic potentials in
the regime of a uniformly positive Lyapunov exponent, that is, for sufficiently large
coupling by a result of Herman. For the particular case of the almost Mathieu operator,
we obtain this result in the regime $\lambda > 2$.

We expect that our method will be useful also in the context of the almost Mathieu
operator at critical coupling, $\lambda = 2$, where almost-diffusive transport is
expected. It is an interesting open problem to bound the transport exponents in this
case. For Diophantine frequencies, Bellissard, Guarneri, and Schulz-Baldes have
established a lower bound on phase-averaged tranport exponents in terms of the
multifractal dimensions of the density of states \cite{bgs}. Since there are no known
lower bounds for these dimensions, however, this result does not yield any explicit
estimates for the dynamical quantities at this point in time.

\end{document}